\documentclass[11pt]{article}

\usepackage[utf8]{inputenc}
\usepackage[T1]{fontenc}
\usepackage{lmodern}           
\usepackage{microtype}       
\usepackage[margin=1in]{geometry}
\usepackage{hyperref}
\usepackage{url}
\usepackage{graphicx}
\usepackage{amsmath,amsfonts,amssymb}
\usepackage{enumitem}
\usepackage[backend=biber,style=ieee]{biblatex}
\usepackage{xcolor}

\title{The AI Fairness Myth: A Position Paper on Context-Aware Bias}

\author{Kessia Nepomuceno, Fabio Petrillo\\
\small École de Technologie Supérieure, Université du Québec\\
\small Montréal, Canada\\
\small\texttt{kessia.cavalcanti-nepomuceno.1@ens.etsmtl.ca, fabio.petrillo@etsmtl.ca}}

\date{} 

\begin{document}
\maketitle

\begingroup
  \renewcommand\thefootnote{}      
  \footnotetext{Disclaimer: This is a position paper; its purpose is to present an idea that has not yet been validated.}
  \addtocounter{footnote}{-1}      
\endgroup

\begin{abstract}
Defining fairness in AI remains a persistent challenge, largely due to its deeply context-dependent nature and the lack of a universal definition. While numerous mathematical formulations of fairness exist, they sometimes conflict with one another and diverge from social, economic, and legal understandings of justice. Traditional quantitative definitions primarily focus on statistical comparisons, but they often fail to simultaneously satisfy multiple fairness constraints. Drawing on philosophical theories (Rawls' Difference Principle and Dworkin’s theory of equality) and empirical evidence supporting affirmative action, we argue that fairness sometimes necessitates deliberate, context-aware preferential treatment of historically marginalized groups. Rather than viewing bias solely as a flaw to eliminate, we propose a framework that embraces corrective, intentional biases to promote genuine equality of opportunity. Our approach involves identifying unfairness, recognizing protected groups/individuals, applying corrective strategies, measuring impact, and iterating improvements. By bridging mathematical precision with ethical and contextual considerations, we advocate for an AI fairness paradigm that goes beyond neutrality to actively advance social justice. 

\end{abstract}

\section{Introduction}\label{sec:intro}

In the literature, it is difficult to find a simple and universal way to reason about the definition of fairness. There is no clear consensus on a single definition that applies to every case. One reason for this is that fairness is inherently context-dependent, it varies depending on the situation in which it is applied or interpreted \cite{kaur2022trustworthy}.
Analyzing different studies, we observe a continuous shift from one fairness metric to another, as researchers attempt to identify the most suitable mathematical definition for specific contexts \cite{caton}. According to Mehrabi \cite{mehrabi2021survey}, the following are the ten most widely used quantitative definitions of fairness: 

(1) Equalized Odds; (2) Equal Opportunity; (3) Demographic Parity; (4) Fairness Through Awareness; (5) Fairness Through Unawareness; (6) Treatment Equality; (7) Test Fairness; (8) Counterfactual Fairness; (9) Fairness in Relational Domains; (10) Conditional Statistical Parity.
\vspace{0.3cm}

Many of these definitions aim to ensure equal opportunities for individuals and groups. With the exception of Definition 9 (Fairness in Relational Domains), most of that definition focus on comparing groups in terms of classification outcomes and probabilities. These approaches often attempt to either compare equality across groups and individuals or remove sensitive attributes to avoid biased decision-making.

Therefore, when working with AI fairness, it is essential to first understand the specific scenario and then adopt the mathematical definition that best fits that context. This approach has guided much of the work on fairness in AI systems. However, a major challenge lies in the limitations of these fairness definitions and the risk of over-relying on metrics.

The multiple mathematical definitions of fairness often fail to align with normative social, economic, or legal understandings. In fact, some quantitative definitions may even contradict one another \cite{mehrabi2021survey, kleinberg2016inherent}.
Liu \cite{liu2018delayed} highlights that current definitions of fairness are not always effective in promoting improvements for sensitive groups. In some cases, they can even be harmful when analyzed over time. Additionally, measurement errors can unintentionally reinforce the appearance of fairness according to certain definitions, while masking underlying issues.

Although we have a wide variety of fairness concepts, each covering different scenarios, their perspectives on what fairness means can often be inconsistent or even conflicting. Synthesizing these diverse definitions into a single one widely applicable, remains a significant challenge. Achieving this would allow for more consistent and comparable evaluations across different systems and contexts. Still, a definition stated in \cite{mehrabi2021survey}, describes fairness as follows: 

\begin{quote}\itshape
``Fairness is the absence of any prejudice or favoritism toward an individual or group based on their inherent or acquired characteristics.''
\end{quote}


Though this definition is widely used, it presents some practical challenges. In real-world systems, achieving a complete absence of prejudice or favoritism is often unrealistic. Enforcing this idea may even distort the system’s context and inadvertently affect fairness. Therefore, it might be more effective to design systems that acknowledge existing prejudices and favoritism and work with them toward reaching a fair decision-making process.

Therefore, our objective with this paper, is to discuss the concepts of fairness by presenting the problems, limitations, and implications associated with its adoption. We also offer a new perspective on how to approach fairness, introducing a context-aware bias view inspired by philosophical and empirical insights. Furthermore, we propose new concept of fairness and, ultimately, present a framework designed to put these ideas into practice.

This paper follows the structure: Section 3 presents our position, offering a new perspective, and a new conceptualization of fairness; Section 4 introduces the proposed framework; Section 5 discusses some counterarguments presented; Section 6 concludes the study.



\section{Position: Centering Advocacy in Fairness}\label{sec:position}

As previously mentioned, although the quoted definition, as well as the mathematical formulation of fairness, is widely used, we argue that it lacks a practical perspective. When we think about fairness, what often comes to mind is an equitable system based on equal treatment. The idea of equal treatment is powerful in theory, it aims for racial and social sameness. However, in practice, this approach assumes that everyone starts from an equal position, which is not the case. People have different histories, backgrounds, and lived experiences.

Therefore, we argue for the necessity of preferential treatment for historically disadvantaged minority groups. In certain contexts, fairness may actually require unequal treatment to ensure justice.
John Rawls, in A Theory of Justice \cite{rawls2017theory}, introduces the Difference Principle, which states that social and economic inequalities must be arranged to benefit the least-advantaged members of society. Similarly, Ronald Dworkin, in What Is Equality? \cite{dworkin2018equality}, argues that when inequalities are the result of luck, such as inherited privilege, the state has a duty to compensate for them, often through direct support or preferences for marginalized groups.

This idea is not just theoretical. The UN General Assembly (1992) \cite{UNGA1992Minorities} and the UN OHCHR (2010) \cite{office2010minority} have both acknowledged the legitimacy of special measures for minorities. They emphasize that states “shall take measures” to create conditions that enable minorities to flourish culturally and participate equally in society.
Empirical evidence supports this approach, a 2023 meta-review by the United Nations University \cite{schotte2023does}, analyzing 194 studies, found that 63\% concluded that affirmative action programs improved educational, employment, and political outcomes for target minorities.
Advocating for marginalized groups, commonly known as positive discrimination, positive action, or affirmative action, is already a practice in many regions.

What we want to emphasize is the dual nature of fairness. On one hand, equal treatment is important. But on the other hand, and perhaps more controversially, it’s crucial to recognize that different contexts may require different approaches, including unequal treatment, to truly advocate for justice.
Bringing this idea into the domain of AI fairness, much of the existing work has focused on equal treatment. However, we suggest a shift in perspective: \emph{Instead of avoiding bias altogether, we should consider how intentional, context-aware biases might actually lead to fairer outcomes.}
\vspace{0.2cm}

The core of our discussion lies in the system’s ability to recognize the differences between groups/individuals, account for them intentionally (considering context-aware biases), and then evaluate whether it ensures equal opportunity. Instead of overlooking the presence of biases, we acknowledge and incorporate them into the process, advocating for affected groups/individuals. Based on this process, we propose a broader conceptualization of fairness goal, one that is grounded in real-world contexts of prejudice and favoritism, yet focused on system outcomes:
\begin{quote}\itshape
Fairness is the principle of ensuring that systems do not perpetuate, reinforce, or amplify existing societal biases.
\end{quote}
This perspective aligns with the principles outlined in NIST Special Publication 1270 \cite{schwartz2022towards}, emphasizing outcome-based fairness while acknowledging the pre-existing nature of societal biases. 

\section{Context-Aware Fairness Framework}\label{sec:workflow}

In this section, we introduce a framework designed to put into practice the concepts discussed thus far. The goal of this framework is not to test the fairness of a system through strict quantitative metrics, but rather to serve as a proof of concept for the ideas presented in this article. So, stepping back from strict measurements, we will dive into a more holistic framework, considering the ethical and contextual dimensions of fairness. Specifically, we aim to demonstrate whether the use of intentional, context-aware biases can actually lead to fairer outcomes.

\begin{itemize}
\item The first step is to \textbf{ensure we are working with an unfair} system. The idea behind starting with an unfair system is to create a biased scenario that can later be evaluated through our framework.

\item Once the unfair system is established, we \textbf{identify the protected and unprotected groups or individuals} within the context. A standard method for this identification can be applied, such as in \cite{chen2019fairness}. The goal of this step is to understand the underlying factors contributing to the unfair outcomes.

\item Next, \textbf{we deliberately take actions to support or uplift the protected groups/individuals} (e.g., through affirmative strategies). These actions should be based by the distinctions between protected and unprotected groups or individuals identified before.

\item Then, we \textbf{introduce context-aware bias,} not as a flaw, but as a deliberate corrective tool to support historically marginalized groups. This step reframes bias as a mechanism for promoting fairness within specific ethical and contextual boundaries.

\item We then \textbf{measure the outcomes} resulting from these interventions.

\item Finally, we \textbf{compare and evaluate} the results obtained with and without the interventions to assess their effectiveness in promoting fairness.
\end{itemize}


This approach emphasizes fairness as not only a statistical construct but also a contextual and ethical one, where equal treatment may sometimes require unequal consideration.

\section{Counterarguments and Discussion}\label{sec:discussion}

In this section, we address potential counterarguments to our proposal, contributing to the broader debate on affirmative action. The first argument concerns whether preferential treatment constitutes reverse discrimination. The second addresses the challenges involved in operationalizing this approach.

\paragraph{Objection 1: Preferential treatment is reverse discrimination.}  
We respond by emphasizing that, without deliberate intervention, algorithms inevitably inherit and often amplify historical and systemic biases embedded in historical data. The notion of algorithmic neutrality is a myth; what appears neutral often reflects and perpetuates existing social inequities. In this context, preferential treatment should not be viewed as an act of favoritism or punishment, but rather as a necessary corrective measure, a form of remedial action intended to counterbalance systemic disadvantages and foster more equitable outcomes.



\paragraph{Objection 2: Hard to operationalize.}  

Operationalizing fairness in AI systems is a challenge, regardless of the fairness measurement or definition being considered \cite{black2023toward}. This difficulty arises because fairness is a multidimensional problem that requires diverse perspectives to fully capture its complexity \cite{nepomuceno2025positionpaper}. Still, we outline a lightweight intervention through a framework that introduces a new perspective on fairness and test it through a proof of concept with minimal overhead.


\section{Conclusion}\label{sec:conclusion}

Defining fairness in AI demands more than selecting a metric from an ever-growing catalogue. Our discussion illustrates three recurring challenges: (1) context-dependence versus universality; (2) mathematical definitions versus what is fair in society; (3) equal versus preferential treatment. Existing quantitative definitions capture narrow statistical regularities, but they struggle to capture the nuances and complexity of fairness, especially when dealing with historical and social inequalities that make “identical treatment” an illusion. 
Drawing on political-philosophical foundations (Rawls, Dworkin) and empirical evidence supporting affirmative action, we argue that fairness sometimes mandates deliberate, context-aware asymmetry to repair historically disadvantage.

So, we suggest thinking about AI fairness by first identifying when a system is unfair or reinforces social bias and then taking actions to fix it, even if that means favoring certain groups until everyone has equal chances. Our proposed workflow (identify groups → apply corrective bias → measure impact → iterate) combines ethics with data to guide fairer outcomes.

Adopting this perspective helps us move beyond the myth that fairness means being “neutral” or completely “unbiased”. Instead, it focuses on building AI systems that expand opportunity for the least advantaged. 
Future research must continue to bridge the gap between mathematical definitions and the broader societal goals of justice, ensuring that AI systems contribute positively to everyone.

\printbibliography
\end{document}